\documentclass[conference]{IEEEtran}

\usepackage[hyphens]{url}
\usepackage{hyperref} 
\usepackage{graphicx}
\usepackage{amsmath}
\usepackage{amssymb}

\usepackage{etoolbox}
\newbool{colored}
\IfFileExists{./colored}{\booltrue{colored}}{\boolfalse{colored}}
\newbool{ascii}
\IfFileExists{./ascii}{\booltrue{ascii}}{\boolfalse{ascii}}

\usepackage{subcaption}
\usepackage{xspace}
\usepackage{booktabs}
\usepackage{enumitem}
\usepackage{numprint}
\usepackage{multirow}
\usepackage[dvipsnames]{xcolor}

\usepackage{tikz}
\usetikzlibrary{positioning,fit,arrows.meta,backgrounds}

\usepackage[frozencache=true,cachedir=.]{minted}
\setminted{fontsize=\footnotesize,baselinestretch=.97,linenos,frame=lines,xleftmargin=6pt,numbersep=3pt,mathescape=true,escapeinside=||,bgcolor=bg}
\definecolor{bg}{rgb}{0.97,0.97,0.97}
\newminted{cpp}{}
\newenvironment{cplus}{\VerbatimEnvironment\begin{cppcode}}{\end{cppcode}}
\newmintinline[cplusinl,mathescape]{cpp}{}

\usepackage[ruled,vlined,linesnumbered]{algorithm2e}

\SetKwProg{Fn}{Function}{}{end}
\SetCommentSty{itshape}

\SetKwComment{Comment}{\color{Green}// }{}

\DontPrintSemicolon
\SetKwRepeat{Do}{do}{while}
\SetKw{KwDownto}{downto}
\SetKw{Continue}{continue}
\SetKw{Break}{break}

\makeatletter
\patchcmd{\ALG@step}{\addtocounter{ALG@line}{1}}{\refstepcounter{ALG@line}}{}{}
\newcommand{\ALG@lineautorefname}{Line}
\makeatother

\makeatletter
\algocf@newcommand{KwDataXX}[1]{%
  \sbox\algocf@inputbox{\hbox{\KwSty{Data}\algocf@typo: }}%
  \ifthenelse{\boolean{algocf@inoutnumbered}}{\relax}{\everypar={\relax}}%
  {\let\\\algocf@newinput\hspace{\wd\algocf@inputbox}\hangindent=\wd\algocf@inputbox\hangafter=\wd\algocf@inputbox#1\par}%
  \algocf@linesnumbered%
}
\makeatother

\newcommand{\ie}{i.e.,\@\xspace}

\newcommand{\eg}{e.g.,\@\xspace}

\newcommand{\wrt}{w.r.t.\@\xspace} %

\newcommand{\grb}{GraphBLAS\xspace}
\newcommand{\ssgrb}{SuiteSparse:GraphBLAS\xspace}

    \newcommand{\grbm}[1]{{\ifbool{colored}{\color{brown}}{}{\mathbf{#1}}}}%
    \newcommand{\grbv}[1]{{\ifbool{colored}{\color{lilac}}{}{\mathbf{#1}}}}%
    \newcommand{\grba}[1]{{\ifbool{colored}{\color{gray}}{}{\textbf{\textit{#1}}}}}%
    \newcommand{\grbs}[1]{{\ifbool{colored}{\color{blue}}{}{\mathit{#1}}}}%

    \newcommand{\grbmask}[1]{\langle #1 \rangle}
    \newcommand{\grbstr}[1]{s(#1)}
    
    \newcommand{\grbmaskreplace}[1]{\langle #1, \mathrm{r} \rangle}
    \newcommand{\grbneg}{\neg}

    \DeclareFontEncoding{LS1}{}{}
    \DeclareFontSubstitution{LS1}{stix}{m}{n}
    \DeclareSymbolFont{arrows1}{LS1}{stixsf}{m}{n}
    \global\let\mapsfrom\undefined %
    \DeclareMathSymbol{\mapsfrom}{\mathrel}{arrows1}{"AB}
    \newcommand{\grbassign}{\mapsfrom}

    \newcommand{\grbf}[2]{\mathit{#1}(#2)}
    \newcommand{\grbreduce}[4]{[ {#1}_{#2}\, #3(#4) ]}
    \newcommand{\grbtransposesymbol}{\mathsf{T}}
    \newcommand{\grbt}{^\grbtransposesymbol} %
    
    \newcommand{\grbdiv}{\grbbinaryop{div}}

    \newcommand{\grbaccum}{\ensuremath{\odot}}
    \newcommand{\grbaccumeq}[1]{\mathbin{\ensuremath{\ifstrempty{#1}{\grbaccum}{#1}\!\!=}}}

    \newcommand{\grbplus}{\oplus}
    \newcommand{\grbtimes}{\otimes}

    \newcommand{\grbbool}{\mathbb{B}}  %
    \newcommand{\grbuint}{\mathtt{UINT64}} %
    \newcommand{\grbfloat}{\mathbb{Q}} %
    \newcommand{\grbdouble}{\mathtt{FP64}} %

    \newcommand{\grbscalartype}[2]{#1_{#2}}
    \newcommand{\grbvectortype}[3]{#1_{#2}^{#3}}
    \newcommand{\grbmatrixtype}[4]{#1_{#2}^{#3 \times #4}}

    \newcommand{\grbnewvector}[4]{\text{let: } #1 \in \grbvectortype{#2}{#3}{#4}}
    \newcommand{\grbnewmatrix}[5]{\text{let: } #1 \in \grbmatrixtype{#2}{#3}{#4}{#5}}

    \newcommand{\grboperator}[1]{\mathsf{#1}}

    \newcommand{\grboperationnoarg}[1]{\mathrm{#1}}

    \newcommand{\grbewiseadd}[1]{\grbbinaryop{#1_\cup}}
    \newcommand{\grbewisemult}[1]{\grbbinaryop{#1_\cap}}

\newcommand{\grbsemiringops}[2]{\mathbin{\grboperator{#1.#2}}}
\newcommand{\grbplustimes}{\grbsemiringops{\grbplus}{\grbtimes}}

\newcommand{\grbpluspair}{\grbsemiringops{plus}{pair}}
\newcommand{\grbplussecond}{\grbsemiringops{plus}{second}}
\newcommand{\grbplusfirst}{\grbsemiringops{plus}{first}}

\newcommand{\grbanysecondi}{\grbsemiringops{any}{secondi}}

\newcommand{\grbminplus}{\grbsemiringops{min}{plus}}

\newcommand{\grbsecondmin}{\grbsemiringops{second}{min}}

\newcommand{\grbarithmeticplustimestext}{\grbsemiringops{plus}{times}}

\newcommand{\grbbinaryop}[1]{\mathbin{\grboperator{#1}}}
\newcommand{\grbany}{\grbbinaryop{any}}
\newcommand{\grbpair}{\grbbinaryop{pair}}

\newcommand{\grbmin}{\grbbinaryop{min}}

\newcommand{\grbfirst}{\grbbinaryop{first}}
\newcommand{\grbsecond}{\grbbinaryop{second}}

\newcommand{\grbsecondi}{\grbbinaryop{secondi}}

\newcommand{\grbarithmeticplus}{\grbbinaryop{+}}

\newcommand{\grbplustext}{\grbbinaryop{plus}}
\newcommand{\grbtimestext}{\grbbinaryop{times}}

\newcommand{\grbgenericop}{\grboperator{op}}

\newcommand{\grbbooleanvalue}[1]{\mathtt{#1}}
\newcommand{\grbtrue}{\grbbooleanvalue{TRUE}}
\newcommand{\grbfalse}{\grbbooleanvalue{FALSE}}

\newcommand{\grbcnt}[1]{| #1 |}

\newcommand{\grboperation}[2]{\grboperationnoarg{#1}(#2)}
\newcommand{\grbnrows}[1]{\grboperation{nrows}{#1}}

\newcommand{\grbnvals}[1]{\grboperation{nvals}{#1}}

\newcommand{\grbselect}[1]{\grbmask{#1}}

\newcommand{\grbtril}[1]{\grboperation{tril}{#1}}
\newcommand{\grbtriu}[1]{\grboperation{triu}{#1}}

\usepackage{microtype}

\usepackage{listings}

\usepackage{textcomp}
\begin{document}

\title{LAGraph: Linear Algebra, Network Analysis Libraries, and the Study of Graph Algorithms}
\author{ %
\IEEEauthorblockN{
    G\'abor Sz\'arnyas\IEEEauthorrefmark{1},
    David A. Bader\IEEEauthorrefmark{2},
    Timothy A. Davis\IEEEauthorrefmark{3},
    James Kitchen\IEEEauthorrefmark{4}, \\
    Timothy G. Mattson\IEEEauthorrefmark{5},
    Scott McMillan\IEEEauthorrefmark{6},
    Erik Welch\IEEEauthorrefmark{4}
} \\
\IEEEauthorblockA{
    \IEEEauthorrefmark{1}CWI Amsterdam
    \IEEEauthorrefmark{2}New Jersey Institute of Technology
    \IEEEauthorrefmark{3}Texas A\&M University
    \IEEEauthorrefmark{4}Anaconda, Inc.
}
\IEEEauthorblockA{
    \IEEEauthorrefmark{5}Intel Corp.
    \IEEEauthorrefmark{6}Software Engineering Institute, Carnegie Mellon University
}
}
\maketitle

\begin{abstract}

Graph algorithms can be expressed in terms of 
linear algebra. GraphBLAS is a library of low-level building blocks for such algorithms
that targets algorithm \emph{developers}.  LAGraph builds on top of the GraphBLAS to
target \emph{users} of graph algorithms with high-level algorithms common in network
analysis.   In this paper, we describe 
the first release of the LAGraph library, the design decisions behind the
library, and performance using the GAP benchmark suite.
LAGraph, however, is much more than a library. It is also a
project to document and analyze the full range of algorithms enabled by the GraphBLAS.  To that end,
we have developed a compact and intuitive notation for describing
these algorithms.  In this paper, we present that 
notation with examples from the GAP benchmark suite.  

\end{abstract}


\section{Introduction}
\label{sec:introduction}

Graphs represent networks of relationships. They play a key role in 
a wide range of applications.   Consequently, numerous graph libraries exist 
such as igraph~\cite{igraph}, NetworkX~\cite{DBLP:reference/snam/X18xv}, and SNAP~\cite{DBLP:journals/tist/LeskovecS16}.
These libraries let programmers work with graphs without the need to master the art of crafting graph algorithms.

There are multiple ways to build libraries of graph algorithms.  One approach
views graphs as sparse matrices and graph algorithms as
linear algebra. This perspective led to the 
GraphBLAS~\cite{DBLP:conf/hpec/MattsonBBBDFFGGHKLLPPRSWY13,DBLP:conf/hpec/MattsonYMBM17}; 
a community effort~\cite{GraphBLASforum} to define low-level building blocks for graph algorithms as linear algebra.
The GraphBLAS are for graph algorithm \emph{developers}.  They are too 
low-level for graph algorithm \emph{users}.  To focus on users and the 
algorithms they require, we launched the
LAGraph project~\cite{DBLP:conf/ipps/MattsonDKBMMY19}.  

LAGraph is a library 
of high quality, production-worthy algorithms constructed on top of
the GraphBLAS.  In this paper, we describe the first release of LAGraph~\cite{LAGraphRepo}.
While LAGraph will eventually work with any implementation of the GraphBLAS, it is currently tied to
the \ssgrb library~\cite{SuiteSparseGraphBLAS} (SS:GrB).

In this release of LAGraph, we restricted ourselves to versions of the algorithms found in the GAP benchmark.
This restricted scope allowed us to focus on the key design decisions needed to establish a solid
foundation for the future.  Those design decisions, the rationale behind them, and a performance baseline 
using the GAP benchmark suite~\cite{DBLP:journals/corr/BeamerAP15} are key contributions of this paper.   

The LAGraph project is more than a library project.   It is also
a repository of algorithms based on the GraphBLAS to help advance the state of the art in 
Graph algorithms expressed as Linear algebra. To support this goal, we created a concise notation for expressing
graph algorithms in terms of the GraphBLAS.   As an example of this notation in action, we use it to describe 
the algorithms used in the GAP benchmark suite.  This notation is a key contribution of this paper.

\section{Design Decisions}
\label{sec:decisions}

LAGraph is for users who want to use graph algorithms that run on top of the GraphBLAS.  
Our overarching design goal is ease of use with flexibility to handle advanced use-cases.
We do not wish to compromise performance, but when the tradeoff between convenience and performance
is unavoidable, we offer both and let the user choose.
LAGraph includes a set of data structures and utility functions 
that make it convenient for developers to write algorithms on top of GraphBLAS with 
an approachable API and consistent user experience.

\begin{listing}
\begin{cplus}
typedef struct LAGraph_Graph_struct
{
    GrB_Matrix   A;      // adjacency matrix of the graph
    LAGraph_Kind kind;   // kind of graph: directed, etc.
    
    // cached properties
    GrB_Matrix   AT;     // transpose of A
    GrB_Vector   row_degree;
    GrB_Vector   col_degree;
    LAGraph_BooleanProperty A_pattern_is_symmetric;
    int64_t      ndiag;  // -1 if unknown
} *LAGraph_Graph;

typedef struct LAGraph_Graph_struct *LAGraph_Graph; |$\label{line:LAGraph_Graph}$|

// creating a graph
GrB_Matrix M;
// ...construction of M omitted

LAGraph_Graph G;
LAGraph_New(&G, &M, LAGRAPH_DIRECTED_ADJACENCY); |$\label{line:CreateGraphObject}$|

// operating on properties
LAGraph_Property_AT(G, msg);  |$\label{line:ComputeTranspose}$|// compute/cache
\end{cplus}
\caption{{\tt LAGraph\_Graph} data structure and methods.}
\label{Lst.graph}
\end{listing}

\subsection{Core data structure}

The main data structure in LAGraph is the \verb'LAGraph_Graph' which consists of primary components
and cached properties. The data structure is not opaque, providing the user with full ability to access and
modify all internal components. This contrasts with the opaque objects in the GraphBLAS.  This data structure is 
shown at the top of Listing~\ref{Lst.graph} and defined ultimately on Line~\ref{line:LAGraph_Graph}.

The primary components of this struct are a GraphBLAS matrix named \verb'A' and an enumeration \verb'kind'.
The kind indicates how the matrix should be interpreted.
Currently, the only kinds defined are \verb'LAGRAPH_ADJACENCY_UNDIRECTED' and 
\verb'LAGRAPH_ADJACENCY_DIRECTED', but more options will be added in the future.  Creating the 
Graph object is performed on Line~\ref{line:CreateGraphObject} of Listing~\ref{Lst.graph}.
Following this call, \verb'M' will be \verb'NULL'. The matrix previously pointed to by \verb'M' now
lives at \verb'G->A'. This ``move'' constructor helps avoid memory-freeing errors.

Cached properties include the transpose of \verb'A', the row degrees, column degrees, etc.
They can be computed from the primary components, but doing so repeatedly for each algorithm
utilizing \verb'A' would be wasteful. Having them live inside the Graph object simplifies
algorithm call signatures. Utility functions exist to compute each cached property.  For example,
Line~\ref{line:ComputeTranspose} of Listing~\ref{Lst.graph} will compute the transpose of \verb'G->A' and store it as \verb'G->AT'.
Following this call, any algorithm which is given \verb'G' will have access to both \verb'A' and its transpose.

Because the Graph object is not opaque, any piece of code may set the transpose as well. For instance, if an algorithm
computes the transpose as part of its normal logic, it could directly set \verb'G->AT'.
The expectation is that the Graph object will always remain consistent.
If \verb'G->A' is modified, all cached properties must be either be set as unknown or modified to reflect the change.
Properties which are not known are set to \verb'NULL' or \verb'LAGRAPH_BOOLEAN_UNKNOWN' in the case of boolean properties.
This expectation is a convention that all LAGraph algorithm implementers are expected to follow.

\subsection{User modes}

Algorithms in LAGraph target two user modes: Basic and Advanced. The Basic user mode is for those who want
things to ``just work'', are less concerned about performance, and may be less experienced with graph libraries.
The Advanced user mode is for those whose primary concern is performance and are willing to conform to stricter
requirements to achieve that goal.

Algorithms targeting the Basic mode typically have limited options. Often, there will only be one function for
a given algorithm. Under the hood, that single algorithm might take different paths depending on the shape or
size of the input graph. The idea is that a basic user wants to compute PageRank or Betweenness Centrality,
but does not want to have to understand the five different ways to compute them. They simply want the correct answer.

Algorithms targeting the Advanced mode are often highly specialized implementations of an algorithm. The Advanced
mode user is expected to understand details such as push-pull~\cite{DBLP:conf/icpp/YangBO18} and batch mode and why different techniques are
better for each graph. Advanced mode algorithms are very strict in their input. If the input does not match the
expected kind, an error will be raised.

Advanced mode algorithms will also raise an error if a cached property is needed by an algorithm, but is not
currently available on the Graph object. While Basic mode algorithms are free to compute and cache properties
on the Graph object, Advanced mode algorithms never will. The idea is to never surprise the user with unexpected
additional computation. An Advanced mode user must opt-in to all computations.

Often, Basic mode algorithms will inspect the input, possibly compute properties or transform the data,
and finally call one of the Advanced mode algorithms to do the actual work on the graph. Having these two user
modes allows LAGraph to target a wider range of users who vary in their experience with graph algorithms.

\subsection{Algorithm calling conventions}

Algorithms in LAGraph follow a general calling convention.

\begin{cplus}
int algorithm
(
    // outputs
    TYPE *out1,
    TYPE *out2,
    ...
    // input/output
    TYPE inout,
    ...
    // inputs
    TYPE input1,
    TYPE input2,
    ...
    // error message holder
    char *msg
)
\end{cplus}

The return value is always an int with the following meaning:

\begin{itemize}
    \item \verb'=0 ->' success
    \item \verb'<0 ->' error
    \item \verb'>0 ->' warning
\end{itemize}

The meaning of a given error or warning value is algorithm-specific
and should be listed in the documentation for the algorithm.

We distinguish three types of arguments:

\begin{itemize}
    \item 
    Outputs appear first and are passed by reference. A pointer should be created by the caller, but
    memory will be allocated by the algorithm. If the output is not needed, a \verb'NULL' is passed and the
    algorithm will not return that output.

    \item Input/Output arguments are passed by value. The expectation is that the object will be modified.
    This supports features such as batch mode in which a frontier is updated and returned to the caller.
    It also supports Basic mode algorithms which may modify a Graph object by adding cached properties.

    \item Inputs are passed by value and should never be changed by the algorithm.
\end{itemize}

The final argument of any LAGraph algorithm holds the error message. This must be \verb'char[]' of
size \verb'LAGRAPH_MSG_LEN'. When the algorithm returns an error or a warning, a message may be placed in
this array as additional information. Because the caller creates this array, the caller must free
the memory or reuse it as appropriate. If the algorithm is successful, it should fill the message array
with an empty string to clear any previous message.

\subsection{Error handling}

Because every algorithm in LAGraph can return an error, the return value of every call should be
checked before proceeding. To make this less burdensome for a C-based library, LAGraph provides a
convenience macro which works similar to try/catch in other languages.

\begin{cplus}
#define LAGraph_TRY(LAGraph_method)
{
    int LAGraph_status = LAGraph_method;
    if (LAGraph_status < 0)
    {
        LAGraph_CATCH (LAGraph_status);
    }
}
\end{cplus}

\verb'LAGraph_CATCH' can be defined before an algorithm and will be called in the event of an error.
This allows for proper freeing of memory and other necessary tasks.

A similar macro, \verb'GrB_TRY', will call \verb'GrB_CATCH' when making GraphBLAS calls which return
a \verb'GrB_Info' value other than \verb'GrB_SUCCESS' or \verb'GrB_NO_VALUE'.

\verb'LAGraph_TRY' and \verb'GrB_TRY' provide an easy to use and easy to read method for dealing with
error checking while writing graph algorithms.

\subsection{Contributing algorithms}

The LAGraph project welcomes contributions from graph practitioners who understand the GraphBLAS vision
of using the language of linear algebra to express graph computations. However, as a matter of
practical concern, many users want a stable experience when using LAGraph for doing real work. To balance
these, the LAGraph repository will have both a stable and an experimental folder.

New algorithms or modifications of existing algorithms will first be added to the experimental folder.
The release schedule of experimental algorithms will generally be much faster than the stable release,
and there is no expectation of a bug-free experience.
The goal is to generate lots of ideas and allow uninhibited contributions to
push the boundary of what is possible with the GraphBLAS. The stable release will be fully tested and
will move much slower, targeting the needs of those who want to use LAGraph as a complete, production-grade
library rather than as a research project.

\section{GraphBLAS Theory and Notation}
\label{sec:notation}

\begin{table*}[htbp]
    \centering
    \begin{tabular}{llr@{}ll}
        \toprule
        \multicolumn{1}{c}{\bf operation/method} & \multicolumn{1}{c}{\bf description}                                                        & \multicolumn{2}{c}{\bf notation}                                                                                                           \\
        \midrule
        \tt mxm                                  & matrix-matrix multiplication                                                               & $\grbm{C} \grbmask{\grbm{M}}        $              & $\grbaccumeq{} \grbm{A} \grbplustimes \grbm{B}$                                       \\
        \tt vxm                                  & vector-matrix multiplication                                                               & $\grbv{w}\grbt \grbmask{\grbv{m}\grbt}   $         & $\grbaccumeq{} \grbv{u}\grbt \grbplustimes \grbm{A}$                                  \\
        \tt mxv                                  & matrix-vector multiplication                                                               & $\grbv{w} \grbmask{\grbv{m}}        $              & $\grbaccumeq{} \grbm{A} \grbplustimes \grbv{u}$                                       \\
        \midrule
        \multirow{2}{*}{\tt eWiseAdd}            & element-wise addition using operator $\grbgenericop$                                       & $\grbm{C} \grbmask{\grbm{M}} $                     & $\grbaccumeq{} \grbm{A} \grbewiseadd{\grbgenericop} \grbm{B}$                         \\
                                                 & on elements in the set union of structures of $\grbm{A}/\grbm{B}$ and $\grbv{u}$/$\grbv{v}$ & $\grbv{w} \grbmask{\grbv{m}} $                     & $\grbaccumeq{} \grbv{u} \grbewiseadd{\grbgenericop} \grbv{v}$                         \\
        \midrule
        \multirow{2}{*}{\tt eWiseMult}           & element-wise multiplication using operator $\grbgenericop$                                 & $\grbm{C} \grbmask{\grbm{M}} $                     & $\grbaccumeq{} \grbm{A} \grbewisemult{\grbgenericop} \grbm{B}$                        \\
                                                 & on elements in the set intersection of structures of $\grbm{A}/\grbm{B}$ and $\grbv{u}$/$\grbv{v}$ & $\grbv{w} \grbmask{\grbv{m}} $                     & $\grbaccumeq{} \grbv{u} \grbewisemult{\grbgenericop} \grbv{v}$                        \\
        \midrule
        \multirow{3}{*}{\tt extract}             & extract submatrix from matrix $\grbm{A}$ using indices $\grba{i}$ and indices $\grba{j}$   & $\grbm{C} \grbmask{\grbm{M}} $                     & $\grbaccumeq{} \grbm{A}(\grba{i}, \grba{j})$                                          \\
                                                 & extract the $\grbs{j}$th column vector from matrix $\grbm{A}$                              & $\grbv{w} \grbmask{\grbv{m}} $                     & $\grbaccumeq{} \grbv{A}(:, \grbs{j})$                                                 \\
                                                 & extract subvector from $\grbv{u}$ using indices $\grba{i}$                                 & $\grbv{w} \grbmask{\grbv{m}} $                     & $\grbaccumeq{} \grbv{u}(\grba{i})$                                                    \\
        \midrule
        \multirow{4}{*}{\tt assign}              & assign matrix to submatrix with mask for $\grbm{C}$                                        & $\grbm{C} \grbmask{\grbm{M}} (\grba{i},\grba{j}) $ & $\grbaccumeq{} \grbm{A}$                                                              \\
                                                 & assign scalar to submatrix with mask for $\grbm{C}$                                        & $\grbm{C} \grbmask{\grbm{M}} (\grba{i},\grba{j}) $ & $\grbaccumeq{} \grbs{s}$                                                                                                                   \\
                                                 & assign vector to subvector with mask for $\grbv{w}$                                        & $\grbv{w} \grbmask{\grbv{m}} (\grba{i}) $          & $\grbaccumeq{} \grbv{u}$                                                                                                                   \\
                                                 & assign scalar to subvector with mask for $\grbv{w}$                                        & $\grbv{w} \grbmask{\grbv{m}} (\grba{i}) $          & $\grbaccumeq{} \grbs{s}$                                                              \\
        \midrule
        \multirow{2}{*}{\tt apply}               & \multirow{2}{*}{apply unary operator $\mathit{f}$ with optional thunk $k$}                 & $\grbm{C} \grbmask{\grbm{M}} $                     & $\grbaccumeq{} \grbf{f}{\grbm{A}, \grbs{k}}$                                          \\
                                                 &                                                                                            & $\grbv{w} \grbmask{\grbv{m}} $                     & $\grbaccumeq{} \grbf{f}{\grbv{u}, \grbs{k}}$                                          \\
        \midrule
        \multirow{2}{*}{\tt select}              & \multirow{2}{*}{apply select operator $\mathit{f}$ with optional thunk $k$}                & $\grbm{C} \grbmask{\grbm{M}} $                     & $\grbaccumeq{} \grbm{A}\grbselect{\grbf{f}{\grbm{A}, \grbs{k}}}$                      \\
                                                 &                                                                                            & $\grbv{w} \grbmask{\grbv{m}} $                     & $\grbaccumeq{} \grbv{u}\grbselect{\grbf{f}{\grbv{u}, \grbs{k}}}$                      \\
        \midrule
        \multirow{3}{*}{\tt reduce}              & row-wise reduce matrix to column vector                                                    & $\grbv{w} \grbmask{\grbv{m}} $                     & $\grbaccumeq{} \grbreduce{\grbplus}{\grbs{j}}{\grbm{A}}{:,\grbs{j}}$                  \\
                                                 & reduce matrix to scalar                                                                    & $\grbs{s} $                                        & $\grbaccumeq{} \grbreduce{\grbplus}{\grbs{i}, \grbs{j}}{\grbm{A}}{\grbs{i},\grbs{j}}$ \\
                                                 & reduce vector to scalar                                                                    & $\grbs{s} $                                        & $\grbaccumeq{} \grbreduce{\grbplus}{\grbs{i}}{\grbm{u}}{\grbs{i}}$                    \\
        \midrule
        \multirow{1}{*}{\tt transpose}           & transpose                                                                                  & $\grbm{C} \grbmask{\grbm{M}} $                     & $\grbaccumeq{} \grbm{A}\grbt$                                                         \\
        \midrule
        \midrule
        \multirow{2}{*}{\tt dup}                 & duplicate matrix                                                                           & $\grbm{C} $                                        & $\grbassign \grbm{A}$                                                                 \\
                                                 & duplicate vector                                                                           & $\grbv{w} $                                        & $\grbassign \grbv{u}$                                                                 \\
        \midrule
        \multirow{2}{*}{\tt build}               & matrix from tuples                                                                         & $\grbm{C}\ $                                       & $\grbassign \left\{ \grba{i}, \grba{j}, \grba{x} \right\} $                           \\
                                                 & vector from tuples                                                                         & $\grbv{w}\ $                                       & $\grbassign \left\{ \grba{i}, \grba{x} \right\} $                                     \\
        \midrule
        \multirow{2}{*}{\tt extractTuples}       & \multirow{2}{*}{extract index arrays ($\grba{i}, \grba{j}$) and value arrays ($\grba{x}$)} & $ \left\{ \grba{i}, \grba{j}, \grba{x} \right\} $  & $\grbassign \grbm{A} $                                                                \\
                                                 &                                                                                            & $ \left\{ \grba{i}, \grba{x} \right\} $            & $\grbassign \grbv{u}   $                                                              \\
        \midrule
        \multirow{2}{*}{\tt extractElement}      & \multirow{2}{*}{extract element to scalar}                                                 & $\grbs{s} $                                        & $\ = \grbm{A}(\grbs{i}, \grbs{j})$                                                    \\
                                                 &                                                                                            & $\grbs{s} $                                        & $\ = \grbv{u}(\grbs{i})$                                                              \\
        \midrule
        \multirow{2}{*}{\tt setElement}          & \multirow{2}{*}{set element}                                                               & $\grbm{C}(\grbs{i}, \grbs{j}) $                    & $\ = \grbs{s}$                                                                        \\
                                                 &                                                                                            & $\grbv{w}(\grbs{i})$                               & $\ = \grbs{s}$                                                                        \\
        \bottomrule
    \end{tabular}
    \caption{GraphBLAS operations and methods based on \cite{DBLP:journals/toms/Davis19,GraphBLASv13}.
        \emph{Notation:}
        Matrices and vectors are typeset in bold, starting with uppercase ($\grbm{A}$) and lowercase ($\grbv{u}$) letters, respectively.
        Scalars including indices are lowercase italic ($\grbs{k}$, $\grbs{i}$, $\grbs{j}$) while arrays are lowercase bold italic ($\grba{x}$, $\grba{i}$, $\grba{j}$).
        $\grbplus$ and $\grbtimes$ are the addition and multiplication operators forming a semiring and default to conventional arithmetic $+$ and $\times$ operators.
        $\grbaccum$ is the accumulator operator.
        Operations can be modified via a descriptor;
        matrices can be transposed ($\grbm{B}\grbt$),
        the mask can be complemented ($\grbm{C}\grbmask{\neg \grbm{M}}$), and
        the mask can be valued (shown above) or structural ($\grbm{C}\grbmask{\grbstr{\grbm{M}}}$).
        A structural mask can also be complemented ($\grbm{C}\grbmask{\grbneg\grbstr{\grbm{M}}}$).
        The result can be cleared (replaced) after using it as input to the mask/accumulator step ($\grbm{C}\grbmaskreplace{\grbm{M}}$).
        Not all methods are listed (creating new operators, monoids, and semirings, clearing a matrix/vector, etc.).
    }
    \label{tab:graphblas-notation}
\end{table*}

In this section, we summarize the key concepts in \grb, then present a concise notation for the operations and methods defined in the \grb standard.
Additionally, we demonstrate how the operations can be interpreted as graph processing primitives if graphs are encoded as adjacency matrices 
and nodes are selected using vectors.\footnote{We use these specialized vectors/matrices here for illustration purposes -- the \grb standard allows the definition of arbitrary vectors/matrices.}

\subsection{Overview}

We first give a brief overview of the theoretical aspects of the \grb. For more details, we refer the reader to tutorials~\cite{gabor_szarnyas_2020_4318870} and the specification documents~\cite{GraphBLASv13,GxBUserGuide}.

\paragraph{Data structures}
\grb builds on the duality between graph and matrix data structures.
Namely, a directed graph $G = (V, E)$ can be represented with a boolean \emph{adjacency matrix} $\grbm{A} \in \grbbool^{|V| \times |V|}$ where $\grbm{A}_{i,j} = \grbtrue$ iff $(v_i, v_j) \in E$.
The adjacency matrices used in \grb algorithms are not necessarily square: \eg induced subgraphs, where source nodes are selected from $V_1 \subseteq V$ and target nodes are selected from $V_2 \subseteq V$, can be represented with $\grbm{A} \in \grbbool^{|V_1| \times |V_2|}$.
The transposition of~$\grbm{A} \in D^{n \times m}$ is denoted with $\grbm{A}\grbt \in D^{m \times n}$ where $\grbm{A}\grbt(i, j) = \grbm{A}(j, i)$.
Compared to $\grbm{A}$, matrix~$\grbm{A}\grbt$ contains the edges in the reverse direction.

\emph{Vectors} can be used to encode data for nodes, \eg $\grbv{u} \in \grbbool^{|V|}$ can be used to select a subset of nodes.
For vectors, $\grbm{u}$ denotes a column vector and $\grbm{u}\grbt$ denotes a row vector.
Vectors and matrices can be defined over different types, \eg an unsigned integer ($\grbuint$) matrix can encode the number of paths between two nodes, while a floating point ($\grbdouble$) matrix can encode edge weights.

In practice, adjacency matrices representing graphs are sparse, \ie most of their elements are \emph{zero}, lending themselves to compressed representations such as CSR/CSC.
The \emph{zero elements} take their values during operations based on the identity of the semiring's $\grbplus$ operation (see below).

\paragraph{Semirings}
\grb uses matrix operations %
to express graph processing primitives, \eg a matrix-vector multiplication $\grbm{A} \grbplustimes \grbm{u}$ finds the incoming neighbors of the set of nodes selected by vector~$\grbv{u}$ in the graph of~$\grbm{A}$.

\grb allows users to perform the multiplication operations over an arbitrary \emph{semiring}.
The multiplication operator $\grbtimes$ is used for combining the values of matching input elements, while the addition operator $\grbplus$ defines how the results should be summarized.
For example, the $\grbminplus$ semiring uses $\grbplustext$ as the multiplication operator to compute the path length and $\grbmin$ as the addition operator to determine the length of the shortest path.
The algorithms presented in this paper use a number of non-conventional semirings such as $\grbanysecondi$, $\grbplusfirst$, and $\grbpluspair$. These are summarized in \autoref{tab:semirings} and defined in \autoref{sec:evaluation}.

\begin{table}
    \centering
    \begin{tabular}{llllr}
        \toprule
        \multicolumn{1}{c}{name} & \multicolumn{1}{c}{$\grbplus$} & \multicolumn{1}{c}{$\grbtimes$} & \multicolumn{1}{c}{$D$} & \multicolumn{1}{c}{zero} \\ \midrule
        conventional             & $\grbplustext$                 & $\grbtimestext$                 & $\grbuint$              & $0$                      \\
        $\grbanysecondi$         & $\grbany$                      & $\grbsecondi$                   & $\grbuint$              & $0$                      \\
        $\grbminplus$            & $\grbmin$                      & $\grbplustext$                  & $\grbdouble$            & $-\infty$                \\
        $\grbplusfirst$          & $\grbplustext$                 & $\grbfirst$                     & $\grbuint$              & $0$                      \\
        $\grbplussecond$         & $\grbplustext$                 & $\grbsecond$                    & $\grbuint$              & $0$                      \\
        $\grbpluspair$           & $\grbplustext$                 & $\grbpair$                      & $\grbuint$              & $0$                      \\
        \bottomrule
    \end{tabular}
    \caption{Semirings used in this paper}
    \label{tab:semirings}
\end{table}

\paragraph{Masks and accumulators}
All \grb operations whose output is a vector or a matrix allow the use of masks to limit the scope of the computation and an accumulator $\grbaccum$, a binary operator, that determines how the result of an operation should be applied to their output. %
The semantics of masks is that the computation should be performed
on a given set of nodes (for vector masks) or
on a given set of edges (for matrix masks).
The accumulator determines how the results should be applied to the (potentially non-empty) output matrix/vector.
The interplay of masks and the accumulators is discussed in the specifications~\cite{GraphBLASv13,GxBUserGuide}.

\paragraph{Notation}
To present our algorithms, we use the mathematical notation given in \autoref{tab:graphblas-notation}.
Matrices and vectors are typeset in bold, starting with uppercase ($\grbm{A}$) and lowercase ($\grbv{u}$) letters, respectively.
Scalars including indices are lowercase italic ($\grbs{k}$, $\grbs{i}$, $\grbs{j}$) while arrays are lowercase bold italic ($\grba{x}$, $\grba{i}$, $\grba{j}$).

\subsection{Operations}
\label{sec:operations}

\paragraph{Matrix multiplication}
\label{sec:mxm}

The \emph{matrix-matrix multiplication} operation $\grbm{A} \grbplustimes \grbm{B}$ expresses a navigation step that starts
in the edges of $\grbm{A}$ and traverses from their endpoints
using the edges of $\grbm{B}$.
The result matrix~$\grbm{C}$ has elements $\grbm{C}_{i,j}$ representing the summarized paths (\eg number of paths, shortest paths) between start node $i$ in the graph of $\grbm{A}$ and end node $j$ in the graph of $\grbm{B}$.

The \emph{vector-matrix multiplication} operation $\grbv{u}\grbt \grbplustimes \grbm{A}$ performs navigation starting from the nodes selected in vector~$\grbv{u}$ along the edges of matrix~$\grbm{A}$.
The result vector~$\grbm{w}$ contains the set of reached nodes with the values computed on the semiring (combining the source node values with the outgoing edge values using $\grbtimes$ then summarizing these for each target node using $\grbplus$).
The \emph{matrix-vector multiplication} operation $\grbm{A} \grbplustimes \grbv{u}$ performs navigation in the reverse direction on the edges of~$\grbm{A}$.

\paragraph{Element-wise addition}

The \emph{element-wise addition} operations
$\grbm{u} \grbewiseadd{\grbgenericop} \grbm{v}$ and
$\grbm{A} \grbewiseadd{\grbgenericop} \grbm{B}$
apply the operator $\grbgenericop$ on the elements selected by the \emph{union of the structures of its inputs},
\ie nodes/edges which are present in at least one of the input matrices.

\paragraph{Element-wise multiplication}

The \emph{element-wise multiplication} operation
$\grbm{u} \grbewisemult{\grbgenericop} \grbm{v}$ and
$\grbm{A} \grbewisemult{\grbgenericop} \grbm{B}$
apply the operator $\grbgenericop$ on the elements selected by the \emph{intersection of the structures of its inputs},
\ie nodes/edges which are present in both inputs.

\paragraph{Extract}
For adjacency matrix~$\grbm{A}$,
the \emph{extract submatrix} operation $\grbm{A}(\grba{i}, \grba{j})$ returns a matrix containing the elements from $\grbm{A}$ with
row indices in $\grba{i}$ and
column indices in $\grba{j}$.
In graph terms, the submatrix represents an induced subgraph where
the source nodes of the edges are in array $\grba{i}$ and
the target nodes of the edges are in array $\grba{j}$.
The \emph{extract vector} operation $\grbv{A}(\grbs{i}, :)$ selects a column vector containing node $\grbs{i}$'s neighbors along incoming edges.
The \emph{extract subvector} operation $\grbv{u}(\grba{i})$ selects the nodes with indices in array $\grba{i}$.

\paragraph{Assign}
The \emph{assign} operation has multiple variants.
The first assigns a matrix to a submatrix selected by row indices $\grba{i}$ and column indices $\grba{j}$:
$\grbm{C} \grbmask{\grbm{M}} (\grba{i},\grba{j}) \grbaccumeq{} \grbm{A}$.
This operator is useful to project an induced subgraph back to the original graph.
The second assigns a vector to a subvector selected by indices $\grba{i}$:
$\grbv{w} \grbmask{\grbv{m}} (\grba{i}) \grbaccumeq{} \grbv{u}$.
Finally, both the selected submatrix/subvector can be assigned with a scalar value:
$\grbm{C} \grbmask{\grbm{M}} (\grba{i},\grba{j}) \grbaccumeq{} \grbs{s}$ and
$\grbv{w} \grbmask{\grbv{m}} (\grba{i}) \grbaccumeq{} \grbs{s}$.
In all cases, the scope of the assignment can be further constrained using masks (see \autoref{sec:masks}).

\paragraph{Apply and select}
The \emph{apply} and \emph{select} operations evaluate a unary operator $f$ with an optional input $k$ (the \emph{thunk}) on all elements of the input matrix/vector. When evaluated on a given element, function $\mathit{f}$ can access the indices of the element, allowing the operation to be constrained on regions of the matrix such as its lower triangle.
In the case of \emph{apply}, denoted with $\grbf{f}{\grbm{A}, \grbs{k}}$ and $\grbf{f}{\grbv{u}, \grbs{k}}$, the resulting elements are returned as part of the output.
The \emph{select} operation requires $f$ to be a boolean function and zeros out elements that return $\grbfalse$.
Intuitively,
$\grbm{A}\grbselect{\grbf{f}{\grbm{A}, \grbs{k}}}$ and $\grbv{u}\grbselect{\grbf{f}{\grbv{u}, \grbs{k}}}$ express filtering on the edges of matrix~$\grbm{A}$ and the nodes of vector~$\grbv{u}$, respectively.

\paragraph{Reduce}
For adjacency matrix~$\grbm{A}$,
the \emph{row-wise reduction} $\grbv{w} \grbmask{\grbv{m}} \grbaccumeq{} \grbreduce{\grbplus}{\grbs{j}}{\grbm{A}}{:,\grbs{j}}$ represents a summarization of the values on outgoing edges for each node (represented by row vector~$\grbm{A}(:, \grbs{j})$) to vector~$\grbv{w}. $ %
For matrix~$\grbm{A}$, the \emph{reduction to scalar} $\grbs{s} \grbaccumeq{} \grbreduce{\grbplus}{\grbs{i}, \grbs{j}}{\grbm{A}}{\grbs{i},\grbs{j}}$ represents a summarization of all edge values.
For vector~$\grbv{u}$, the \emph{reduction to scalar} $\grbs{s} \grbaccumeq{} \grbreduce{\grbplus}{\grbs{i}}{\grbm{u}}{\grbs{i}}$ represents a summarization of all node values.

\paragraph{Transposition}
Transposition can be applied as a standalone \grb operation $\grbm{C} \grbmask{\grbm{M}} \grbaccumeq{} \grbm{A}\grbt$ %
and also to the input/output matrices of operations, for example:
$$\grbm{C}^{[\grbtransposesymbol]} \grbmask{\grbm{M}} \grbaccumeq{} \grbm{A}^{[\grbtransposesymbol]} \grbplustimes \grbm{B}^{[\grbtransposesymbol]}$$

\subsection{Masks}
\label{sec:masks}

Masks are used  to limit the scope of \grb operations \wrt their outputs.
For operations resulting in a vector, the mask is based on a vector~$\grbm{m}$. For those resulting in a matrix, it is based on a matrix~$\grbm{M}$.
Here, we only discuss matrix masks.  Extension to vectors is straightforward.

By default, the elements of the mask that exist and are non-zero select corresponding elements of the output matrix that should be computed.  
There are three variations on the mask that impact the output of a \grb operation:
\begin{enumerate}
    \item
        Does the computation need to be performed on the elements selected by the mask ($\grbmask{\grbm{M}}$) or the complement of these elements ($\grbmask{\grbneg\grbm{M}}$)?
    \item
        How are existing elements of the output matrix treated that fall outside the ones selected by the  mask?
        By default, masks use \emph{merge} semantics, \ie the computation can only affect elements selected by the mask, elements outside the mask are unaffected.
        If \emph{replace} semantics is set, masks annihilate all elements outside the mask. This is denoted with $\grbmaskreplace{\grbm{M}}$.
    \item
        How the elements are selected?
        By default, masks are \emph{valued}, \ie values in the mask are checked and elements with explicit zero values (\eg 0 for $\grbarithmeticplustimestext$) are not considered to be part of the mask.
        To only consider the pattern of the mask, \ie the elements of the mask that exist, a \emph{structural mask} should be used, denoted with $\grbmask{\grbstr{\grbm{M}}}$. %
\end{enumerate}

Operations can use \emph{replace semantics} and \emph{structural masks} at the same time, denoted with
$\grbmaskreplace{\grbstr{\grbm{M}}}$

\subsection{Methods}

\grb provides methods for initializing and duplicating vectors and matrices
(\eg $\grbnewvector{\grbv{w}}{\grbfloat}{32}{n}$ and $\grbm{C} \grbassign \grbm{A}$),
setting the values of individual elements ($\grbv{w}(2) = \grbtrue$),
extracting the tuples in the form of index/value arrays from matrices/vectors and building them from tuples ($\grbv{w} \grbassign \left\{ \grba{i}, \grba{x} \right\}$ and $\left\{ \grba{i}, \grba{x} \right\} \grbassign \grbv{u}$).
Additionally, methods are provided for creating new operators, monoids, and semirings, clearing a matrix/vector, etc.

\section{Algorithms}
\label{sec:algorithms}

\subsection{Breadth-First Search (BFS)}
\label{sec:bfs}

The breadth-first search (BFS)
builds on the observation that vector-matrix multiplication $\grbv{f}\grbt\grbm{A}$ expresses 
navigation from the nodes selected by vector $\grbv{f}$ in the graph represented
by $\grbm{A}$.
The direction-optimizing push/pull BFS \cite{DBLP:conf/sc/BeamerAP12} is simple
to express in GraphBLAS \cite{DBLP:conf/icpp/YangBO18}.  If $\grbm{A}$ is held by row,
then $\grbv{f}\grbt\grbm{A}$ is a push step, while $\grbm{B}\grbv{f}$ is a pull step, where
$\grbm{B}=\grbm{A}\grbt$ is the explicit transpose of $\grbm{A}$, also held by row.
Other \grb libraries, \eg GraphBLAST, store both directions and perform
direction-optimization automatically~\cite{DBLP:journals/corr/abs-1908-01407}.
The push-only BFS is shown in
\autoref{alg:bfs-parents}, while the push/pull BFS is \autoref{alg:bfs-parents-do}.

The GraphBLAS BFS relies on the $\grbanysecondi$ %
semiring to compute a single step,
$\grbv{q}\grbt \grbmaskreplace{\grbneg \grbstr{\grbv{p}\grbt}}  = \grbv{q}\grbt\grbm{A}$, where $\grbv{q}$ is the current frontier,
$\grbv{p}$ is the parent vector, and $\grbm{A}$ is the adjacency matrix.
The mask is a \emph{complemented structural mask} which means the mask corresponds to the 
empty elements of the mask vector.   Replace semantics are indicated (due to the \emph{r} in the mask expression)
so any elements of the vector other than those selected by the mask are deleted.
The result is the assignment to the parent vector on line 8 updates the vector with the parents of the the newly visited nodes.

Consider a matrix multiply for conventional linear algebra, where the $\grbplus$ %
monoid sums a set of $t$ entries to obtain a single scalar for computing
$c_{ij} = \sum a_{ik} b_{kj}$ in the matrix multiply $\grbm{C} = \grbm{A}\grbm{B}$.  The $\grbany$ %
monoid performs the reduction of $t$ entries to a single number by merely selecting
any one of the $t$ entries as the result $c_{ij}$.  The selection is done
non-deterministically, allowing for a benign race condition.  In the BFS, this
corresponds to selecting any valid parent of a newly discovered node.  Indeed,
the creation of the $\grbany$ %
operator was inspired by Scott Beamer's \verb'bfs.cc'
method in the GAP benchmark, which has the same benign race condition.  The $\grbany$ %
monoid translates the concept of this benign race condition to construct a
valid BFS tree into a linear algebraic operation, suitable for implementation
in GraphBLAS.

The $\grbsecondi$ %
operator is the multiplicative operator in the $\grbanysecondi$ %
semiring, where the result of $a_{ik} b_{kj}$ is simply the index $k$ in the
semiring for $\grbm{C} = \grbm{A}\grbm{B}$.  This gives the id of the parent node for a newly
discovered node in the next frontier.  The $\grbany$ %
monoid then selects any valid
parent $k$.
\begin{algorithm}[htb]
    \caption{Parents BFS (push-only).}
    \label{alg:bfs-parents}
    \DontPrintSemicolon
    \KwIn{$\grbm{A}, \grbs{startVertex}$}
    \Fn{ParentsBFS}{
        $\grbv{p}(\grbs{startVertex}) = \grbs{startVertex}$ \;
        $\grbv{q}(\grbs{startVertex}) = \grbs{startVertex}$ \;
        \For{$\grbs{level} = 1$ \KwTo $\grbnrows{\grbm{A}}-1$}{
            $\grbv{q}\grbt \grbmaskreplace{\grbneg \grbstr{\grbv{p}\grbt}} = \grbv{q}\grbt \grbanysecondi \grbm{A}$ \;
            $\grbv{p} \grbmask{\grbstr{\grbv{q}}} = \grbv{q}$ \;
            \If{$\grbnvals{\grbv{q}} = 0$}{return}
        }
    }
\end{algorithm}

\begin{algorithm}[htb]
    \caption{Direction-Optimizing Parent BFS.}
    \label{alg:bfs-parents-do}
    \DontPrintSemicolon
    \KwIn{$\grbm{A}, \grbm{A}\grbt, \grbs{startVertex}$}
    \Fn{DirectionOptimizingBFS}{
        $\grbv{q}(\grbs{startVertex}) = 0$ \;
        \For{$\grbs{level} = 1$ \KwTo $\grbnrows{\grbm{A}}-1$}{
            \If{$\mathit{Push}(\grbm{A}, \grbv{q})$}{ %
                $\grbv{q}\grbt \grbmaskreplace{\grbneg \grbstr{\grbv{p}\grbt}} = \grbv{q}\grbt \grbanysecondi \grbm{A}$
            }
            \Else{
                $\grbv{q} \grbmaskreplace{\grbneg \grbstr{\grbv{p}}} = \grbm{A}\grbt \grbanysecondi \grbv{q}$
            }
            $\grbv{p} \grbmask{\grbstr{\grbv{q}}} = \grbv{q}$ \;
            \If{$\grbnvals{\grbv{q}} = 0$}{return}
        }
    }
\end{algorithm}

\subsection{Betweenness Centrality (BC)}
\label{sec:bc}

\begin{algorithm}[htb]
	\caption{Betweenness centrality.}
	\label{alg:bc}
	\Fn{BrandesBC}{
		\Comment{$\grbm{P}(k,j)$ = \# paths from $k$th source to node $j$}
		\Comment{$\grbm{F}$: \# paths in the current frontier}
		$\grbnewmatrix{\grbm{P}}{\grbfloat}{64}{\grbs{ns}}{\grbs{n}}$ \;
		$\grbnewmatrix{\grbm{F}}{\grbfloat}{64}{\grbs{ns}}{\grbs{n}}$ \;
                $\grbm{P}(1:k, \grba{s}) = 1$ \;
                \Comment{First frontier:}
                $\grbm{F} \grbmask{\grbneg\grbstr{\grbm{P}}} =\grbm{P} \grbplusfirst \grbm{A}$  \;

		\Comment{BFS phase:}
		\For{$\grbs{d} = 0$ \KwTo $\grbnrows{\grbm{A}}$}{
			$\grbnewmatrix{\grbm{S}[\grbs{d}]}{\grbbool}{}{\grbs{ns}}{\grbs{n}} $ \;
			$\grbm{S}[\grbs{d}]\grbmask{\grbstr{\grbm{F}}} = 1$ \Comment{$\grbm{S}[d]$ = pattern of $\grbm{F}$}
			$\grbm{P} \grbaccumeq{+} \grbm{F}$ \;
			$\grbm{F}\grbmaskreplace{\grbneg\grbstr{\grbm{P}}} = \grbm{F} \grbplusfirst \grbm{A}$ \;
                        \If{$\grbnvals{\grbm{F}} = 0$}{break}
		}

		\Comment{Backtrack phase:}
		$\grbnewmatrix{\grbm{B}}{\grbfloat}{64}{\grbs{ns}}{\grbs{n}}$ \;
		$\grbm{B}(:) = 1.0$ \;
		$\grbnewmatrix{\grbm{W}}{\grbfloat}{64}{\grbs{ns}}{\grbs{n}}$ \;

		\For{$\grbs{i} = \grbs{d} - 1$ \KwDownto $0$}{
			$\grbm{W}\grbmaskreplace{\grbstr{\grbm{S}[\grbs{i}]}} = \grbm{B}  \grbdiv_\cap \grbm{P}$ \;
			$\grbm{W}\grbmaskreplace{\grbstr{\grbm{S}[\grbs{i} - 1]}} = \grbm{W} \grbplusfirst \grbm{A}\grbt$ \;
			$\grbm{B} \grbaccumeq{+} \grbm{W} \times_\cap \grbm{P}$ \;
		}

		\Comment{$\grbv{centrality}(j) = \sum_i (\grbm{B}(i,j) - 1)$}
		$\grbv{centrality}(:) = -\grbs{ns}$ \;
		$\grbv{centrality} \grbaccumeq{+} \grbreduce{+}{\grbs{i}}{\grbm{B}}{\grbs{i},:} $ \;
	}
\end{algorithm}

The vertex betweenness centrality metric is based on the number of
shortest paths through any given node,
$ \sum_{s \ne i \ne t} {\sigma (s, t|i)}/{\sigma(s,t)}, $
where $\sigma(s,t)$ is the total number of shortest paths from node $s$ to $t$,
and $\sigma(s,t|i)$ is the total number of shortest paths from node $s$ to $t$
that pass through node $i$.  This is expensive to compute, so in practice,
a subset of source nodes are chosen at random (a {\em batch}), of size $\mathit{ns}$.

Like the BFS, direction-optimization is incredibly simple to add to the LAGraph
algorithm for batched betweenness centrality (BC).
It only requires a simple heuristic to determine which
direction to use, followed by masked matrix-matrix multiplication with the
matrix or its transpose: $\grbm{F} \grbmask{\grbneg \grbstr{\grbm{P}}} = \grbm{F}\grbm{B}\grbt$ (the pull) or $\grbm{F}
\grbmask{\grbneg \grbstr{\grbm{P}}} = \grbm{F} \grbm{A}$ (the push), where $\grbm{A}$ is the adjacency matrix of
the graph and $\grbm{B} = \grbm{A}\grbt$ is its explicit transpose, $\grbm{F}$ is the frontier, and the
complemented structural mask $\grbneg \grbstr{\grbm{P}}$ is the set of unvisited nodes.  The multiplication
$\grbm{F} \grbm{B}\grbt$ relies on the descriptor to represent the transpose of $\grbm{B}$, which is not
explicitly transposed.  In the backward phase, the pull step is $\grbm{W} = \grbm{W} \grbm{A}\grbt$ while
the push is $\grbm{W} = \grbm{W} \grbm{B}$, where $\grbm{W}$ is the $\mathit{ns}$-by-$n$ matrix in which centrality is
accumulated (where $\mathit{ns}=4$ is a typical batch size).

To simplify the presentation of the entire BC algorithm, \autoref{alg:bc} does
not show the direction-optimization.  It is the same transformation as the pair
of BFS algorithms, where the push-only step (line 5 of
\autoref{alg:bfs-parents}), is expanded to a push/pull heuristic (lines 4-7 of
\autoref{alg:bfs-parents-do}).

\subsection{PageRank (PR)}
\label{sec:pagerank}

PageRank (PR) computes the importance of each node as a recursively-defined
metric: a web page is important if important pages link to it.
\autoref{alg:pagerank} shows the GraphBLAS implementation of PR as specified in
the GAP benchmark.  It uses the $\grbplussecond$ semiring, where
$\grboperator{second}(x,y)=y$, so it can ignore any edge weights in the input
matrix.  The PR in GAP does not properly handle dangling vertices in the graph.
The Graphalytics benchmark has a PageRank variant which avoids this
problem~\cite{DBLP:journals/corr/abs-2011-15028}.  We have included this
version to compare its performance with the GAP benchmark algorithm
\verb'pr.cc'.
\begin{algorithm}[htb]
    \caption{PageRank (as specified in the GAPBS).}
    \label{alg:pagerank}
    \KwData{$\grbm{A} \in \grbbool^{n \times n}$ \Comment*{adjacency matrix}}
    \KwDataXX{$\grbs{damping}$ \Comment*{damping factor}}
    \KwDataXX{$\grbs{tol}$ \Comment*{stopping tolerance}}
    \KwDataXX{$\grbs{itermax}$\Comment*{maximum number of iterations}}
    \KwResult{$\grbv{r} \in \grbfloat^n$}
    \Fn{PageRank}{

        $\grbs{teleport} = \frac{1 - \alpha}{n}$ \;
        $\grbv{r}(0:n-1) = \frac{1}{n}$, $\grbv{t} = \grbfloat^n$ \;
        $\grbv{d_{out}} = \grbreduce{+}{j}{\grbm{A}}{:, j}$ \Comment{precomputed rowdegree}
        $\grbv{d} = \grbv{d_{out}} \grbewisemult{\grbdiv} \grbs{damping}  $ \Comment{prescale with damping}

        \For{$\grbs{k} = 1$ \KwTo $\grbs{numIterations}$}{
            swap $\grbv{t}$ and $\grbv{r}$ \Comment{$\grbv{t}$ is now the prior rank}
            $\grbv{w} = \grbv{t} \grbewisemult{\grbdiv} \grbv{d}$ \;
            $\grbv{r}(0:n-1) = \grbs{teleport}$ \;
            $\grbv{r} \grbaccumeq{+} \grbm{A}\grbt \grbplussecond \grbv{w}$ \;
            $\grbv{t} \grbaccumeq{-} \grbv{r}$ \;
            $\grbv{t} = abs(\grbv{t})$ \;
            \If{$\grbreduce{+}{i}{\grbv{t}}{i} < \grbs{tol}$}{
                return \Comment{since 1-norm of change is small}
            }
        }
    }
\end{algorithm}

\subsection{Single-Source Shortest Paths (SSSP)}
\label{sec:sssp}

A Delta-Stepping Single-Source Shortest Paths algorithm in GraphBLAS is shown in
\autoref{alg:sssp-delta-stepping}.  It relies on the $\grbminplus$ semiring.
Since it is a fairly complex algorithm, refer to
\cite{DBLP:conf/ipps/SridharBMSLM19} for a description of the method.
\begin{algorithm}[htb]
	\caption{SSSP (delta-stepping).}
	\label{alg:sssp-delta-stepping}
	\KwData{\;
		$\quad \grbm{A}, \grbm{A_H}, \grbm{A_L} \in \grbmatrixtype{\grbfloat}{}{\grbcnt{V}}{\grbcnt{V}} $ \;
		$\quad \grbs{s}, \grbs{i} \in \grbscalartype{\grbuint}{} $ \;
		$\quad \Delta \in \grbscalartype{\grbfloat}{} $ \;
		$\quad \grbv{t}, \grbv{t_{Req}} \in \grbvectortype{\grbfloat}{}{\grbcnt{V}} $ \;
		$\quad \grbv{t_{B_i}}, \grbv{e} \in \grbvectortype{\grbuint}{}{\grbcnt{V}} $ \; %
	}
	\Fn{DeltaStepping}{
		$\grbm{A_L} = \grbm{A}\grbselect{0 < \grbm{A} \leq \Delta} $ \;
		$\grbm{A_H} = \grbm{A}\grbselect{\Delta < \grbm{A}} $ \;
		$\grbv{t}(:) = \infty $ \;
		$\grbv{t}(\grbs{s}) = 0 $ \;
		\While{$\grbnvals{ \grbv{t}\grbselect{\grbs{i} \Delta \leq \grbv{t}} } \neq 0$}{
			$\grbs{s} = 0 $ \;
			$\grbv{t_{B_i}} = \grbv{t} \grbselect{\grbs{i} \Delta \leq \grbv{t} < (\grbs{i} + 1) \Delta}$ \;
			\While{$\grbv{t_{B_i}} \neq 0$}{
				$\grbv{t_{Req}} = \grbv{t} \grbewisemult{\times} \grbv{t_{B_i}}$ \;
				$\grbv{t_{Req}} = \grbm{A_L\grbt} \grbminplus \grbv{t_{Req}}$ \;
				$\grbv{e} = \grbv{t} \grbselect{0 < \grbv{e} \grbplus \grbv{t_{B_i}} }$ \;
				$\grbv{t_{B_i}} = \grbv{t} \grbselect{\grbs{i} \Delta \leq \grbv{t_{Req}} < (\grbs{i} + 1) \Delta}$ \;
				$\grbv{t_{B_i}} = \grbv{t_{B_i}} \grbselect{\grbv{t_{Req}} < \grbs{t}}$ \;
				$\grbv{t} = \grbv{t} \grbewiseadd{\grbmin} \grbv{t_{Req}}$ \;
			}
			$\grbv{t_{Req}} = \grbm{A_H\grbt} \grbminplus (\grbv{t} \grbewisemult{\times} \grbv{e})$ \;
			$\grbv{t} = \grbv{t} \grbewiseadd{\grbmin} \grbv{t_{Req}}$ \;
			$\grbs{i} = \grbs{i} + 1 $ \;
	}
	}
\end{algorithm}

\subsection{Triangle Counting (TC)}
\label{sec:triangle-count}

The triangle counting (TC) problem is to compute the number of unique cliques
of size 3 in a graph.  The TC algorithm is shown in
\autoref{alg:triangle-count-sandiadot}, based on \cite{8091043}.
\begin{algorithm}[htb]
	\caption{Triangle counting.}
	\label{alg:triangle-count-sandiadot}
	\KwData{$\grbm{A} \in \grbbool^{n \times n}$}
	\KwResult{$t \in \grbuint$}
	\Fn{TriangleCount}{
                sample the $\grbs{mean}$ and $\grbs{median}$ degree of $\grbm{A}$ \;
                \If{$\grbs{mean} > 4 \times \grbs{median}$}{
                    $\grbv{p}$ = permutation to sort degree, ascending order \;
                    $\grbm{A} = \grbm{A}(\grbv{p},\grbv{p})$ \;
                }
		$\grbm{L} = \grbtril{\grbm{A}}$ \;
		$\grbm{U} = \grbtriu{\grbm{A}}$ \;
		$\grbm{C}\grbmask{\grbstr{\grbm{L}}} = \grbm{L} \grbpluspair \grbm{U}\grbt$ \;
		$\grbs{t} = \grbreduce{\grbarithmeticplus}{\grbs{i}\grbs{j}}{\grbm{C}}{\grbs{i}, \grbs{j}}$ \;
	}
\end{algorithm}

It starts with a heuristic that decides when
to sort the input graph by ascending degree.  Next, it constructs the lower and
upper triangular part and computes a masked matrix multiply using the
$\grbpluspair$ semiring.  Internally, a dot product method is used in SS:GrB,
because $\grbm{U}$ is transposed via the descriptor.  The $\grboperator{pair}$
is the simple function $\grboperator{pair}(x,y)=1$.  When used in a semiring,
it acts like the $\grboperator{times}$ operator of the conventional semiring,
except that it can ignore the values of its inputs and treat them both as~1.
This semiring is useful for structural computations, such as triangle counting,
when the edge weights of a graph may be present but should be ignored in a
particular algorithm.

\subsection{Connected Components}
\label{sec:connected-components}

The connected components algorithm in LAGraph (\autoref{alg:fastsv})
is written by Zhang, Azad, and Bulu{\c{c}}
\cite{ZHANG202014,DBLP:conf/ppsc/ZhangAH20}.  The method maintains a forest of
trees represented by a parent vector, and iteratively merges trees until no
more merging is possible.  The method as shown in \autoref{alg:fastsv} is a
simplified variant that operates on the entire graph.  In the LAGraph
version, a subgraph is constructed first, and the method finds the connected
components of the subgraph, and then operates on the entire graph.
\begin{algorithm}[htb]
	\caption{Connected components (FastSV).}
	\label{alg:fastsv}
	\Fn{FastSV}{
        $\grbs{n} = \grbnrows{\grbm{A}}$ \;
        $\grbv{gf} = \grbv{f}$ \;
        $\grbv{dup} = \grbv{gf}$ \;
        $\grbv{mngf} = \grbv{gf}$ \;
        $\{ \grba{i}, \grba{x} \} \grbassign \grbv{f}$ \;
        \Repeat{$\grbs{sum} == 0$}{
            \Comment{Step 1: Stochastic hooking}
            $\grbv{mngf} = \grbv{mngf} \grbmin \grbm{A} $ \;
            $\grbv{mngf} = \grbv{mngf} \grbsecondmin \grbv{gf}$ \;
            $\grbv{f}(\grba{x}) = \grbv{f} \grbmin \grbv{mngf} $ \;
            \Comment{Step 2: Aggressive hooking}
            $\grbv{f} = \grbv{f} \grbmin \grbv{mngf} $ \;
            \Comment{Step 3: Shortcutting}
            $\grbv{f} = \grbv{f} \grbmin \grbv{gf} $ \;
            \Comment{Step 4: Calculate grandparents}
            $\{ \grba{i},  \grba{x} \} \grbassign \grbv{f}$ \;
            $\grbv{gf} = \grbv{f}(\grba{x})$ \;
            \Comment{Step 5: Check termination}
            $\grbv{diff} = \grbv{dup} \neq \grbv{gf} $ \; %
            $\grbs{sum} = [+_{\grbs{i}} \grbv{diff}(\grbs{i}) ] $ \;
            $\grbv{dup} = \grbv{gf}$ \; %
        }
	}
\end{algorithm}

\section{Utility Fuctions}
\label{sec:utility}

LAGraph includes a set of utility functions that operate
on a graph.  All function names are prefixed with \verb'LAGraph_'
so we exclude that prefix in the names below, for brevity.

\begin{itemize}

\item {\bf Graph Properties:}
    An \verb'LAGraph_Graph' includes cached properties which can be
    assigned by Basic methods, or which are required by Advanced methods.

      \verb'DeleteProperties' clears all properties,
      \verb'Property_AT' computes the transpose of the adjacency matrix \verb'G->A',
      \verb'Property_RowDegree' computes the row degrees of \verb'G->A',
      \verb'Property_ColDegree' computes the column degrees of \verb'G->A',
      and
      \verb'Property_ASymmetricPattern' determines if the pattern of \verb'G->A' is symmetric or unsymmetric.

\item {\bf Display and debug:}
    \verb'CheckGraph' checks the validity of a graph.
    Since the graph is not opaque, a user application is able to change a graph
    arbitrarily and thus might make it an invalid object.
    \verb'DisplayGraph' displays a graph and its properties.

\item {\bf Memory management:}
    Wrappers for \verb'malloc', \verb'calloc', \verb'realloc', and \verb'free' are provided,
    allowing a user application to select the memory manager to be used.
    These default to the ANSI C11 library functions.

\item {\bf Graph I/O:}
    \verb'BinRead' and \verb'BinWrite' read/write a \verb'GrB_Matrix' in binary form.
    \verb'MMRead' and \verb'MMWrite' read/write a \verb'GrB_Matrix' in Matrix Market form.

\item {\bf Matrix operations:}
    \verb'Pattern' returns a boolean matrix containing the pattern of a matrix.
    \verb'IsEqual' determines if two matrices are equal.  It selects the appropriate
    \verb'GrB_EQ_T' operator that matches the matrix type, and then calls \verb'IsAll'.
    \verb'IsAll' compares two matrices and returns false if
    the pattern of the two matrices differ.  It then uses a given comparator operator to
    compare all pairs of entries, and returns true if all comparisons return true.

\item {\bf Degree operations:}
    \verb'SortByDegree' returns a permutation that sorts a graph by its row/column degrees, and
    \verb'SampleDegree' computes a quick estimate of the mean and median row/column degrees.

\item {\bf Error handling:}
    \verb'LAGraph_TRY' and \verb'GrB_TRY' are helper macros for a simple try/catch
    mechanism.  They require the user application to define \verb'LAGraph_CATCH'
    and \verb'GrB_CATCH'.

\item {\bf Other:}
    \verb'TypeName' returns a string with the name of a \verb'GrB_Type'.
    \verb'KindName' returns a string with of graph kind (directed or undirected).
    \verb'Tic' and \verb'Toc' provide a portable timer.
    \verb'Sort1', \verb'Sort2', and \verb'Sort3' sort 1, 2, or 3 integer arrays.

\end{itemize}

\section{Evaluation}
\label{sec:evaluation}

The performance of LAGraph can only be considered in context of an
implementation of the underlying GraphBLAS library.  This is discussed in
Section~\ref{sec:extensions}, followed by performance results of the new
LAGraph API on the 6 algorithms in the GAP Benchmark
\cite{DBLP:conf/sc/BeamerAP12}.

\subsection{SuiteSparse Extensions}
\label{sec:extensions}

In a prior paper (\cite{DBLP:conf/iiswc/AzadABBCDDDDFGG20}), an early draft of
SS:GrB, (SuiteSparse:GraphBLAS v4.0.0, Aug 2020), was compared with the GAP benchmark
\cite{DBLP:conf/sc/BeamerAP12} and four other graph libraries.  This prior
version of SS:GrB included two primary data structures for its sparse matrices:
compressed sparse vector, and a hypersparse variant
\cite{DBLP:conf/ipps/BulucG08}, both held by row or by column.  It included a
draft implementation of a bitmap data structure that could only be used in a
prototype breadth-first search.  Since then, SuiteSparse:GraphBLAS v4.0.3 has
been released, with full support for bitmap and full matrices for all its
operations.  In an $m$-by-$n$ bitmap matrix, the values are held in a full
array of size $mn$, and another \verb'int8_t' array of size $mn$ holds the
sparsity pattern of the matrix.  A full matrix is a simple dense array of size
$mn$.

The bitmap format is particularly important for the ``pull'' phase of an
algorithm, as used in direction-optimizing breadth-first-search
\cite{DBLP:conf/sc/BeamerAP12,DBLP:conf/icpp/YangBO18}.  The GAP benchmark suite uses this method by
holding its frontier as a bitmap in the pull step and as a list in the push
step. The GAP BFS was typically the fastest BFS amongst the 6 graph
libraries compared in \cite{DBLP:conf/iiswc/AzadABBCDDDDFGG20} (for 4 of the 5
benchmark graphs).  With the addition of the bitmap format to SS:GrB,
LAGraph+SS:GrB is able to come within a factor of 2 or so of the performance of
the highly-tuned BFS GAP benchmark (see the results in the next section), for
those 4 graphs.  At the same time, however, the BFS is very easily expressed in
LAGraph as easy-to-read and easy-to-write code.  This enables non-experts to
obtain a reasonably high level of performance with modest programming effort
when writing graph algorithms.

Additional optimizations added to SS:GrB in the past year include a {\em lazy
sort}.  Normally, SS:GrB keeps its vectors sorted (row vectors in a CSR matrix,
or column vectors if the matrix is held by column), with entries sorted in
ascending order of column or row index, respectively.  This simplifies \
algorithms that operate on a \verb'GrB_Matrix'.  However, some algorithms
naturally produce a jumbled result (matrix multiply in particular), while others
are tolerant of jumbled input matrices.  We thus allow the sort to
be left pending.  The lazy sort joins two other kinds of pending work in
SS:GrB: {\em pending tuples} and {\em zombies}~\cite{DBLP:journals/toms/Davis19}.
A pending tuple is an entry
that is held inside a matrix in an unsorted list, awaiting insertion into the
CSR/CSC format of a \verb'GrB_Matrix'.  A zombie is the opposite: it is an
entry in the CSR/CSC format that has been marked for deletion, but has not yet
been deleted from the matrix.  With  lazy sort, the sort is postponed until
another algorithm requires sorted input matrices.  If the sort is lazy enough,
it might never occur, which is the case for the LAGraph BFS and BC.

Positional binary operators have also been added,
such as the $\grbanysecondi$ semiring, %
which makes the BFS much faster.

\subsection{Performance Results}

We ran our benchmarks on an NVIDIA DGX Station (donated to Texas A\&M by
NVIDIA).  It includes a 20-core Intel(R) Xeon(R)
CPU E5-2698 v4 @ 2.20GHz, with 40 threads.  All codes were compiled with gcc
5.4.0 (-O3).  All default settings were used, which means  hyperthreading
was enabled.  The system has 256GB of RAM in a single socket.
LAGraph (Feb 15, 2021) and SuiteSparse:GraphBLAS 4.0.4-draft (Feb 15, 2021) were
used.  The NVIDIA DGX Station includes four P100 GPUs, but no GPUs were used by
this experiment (a GPU-accelerated SS:GrB is in progress).
Table~\ref{table:results} lists the run time (in seconds) for the GAP benchmark
and LAGraph+SS:GrB for the 6 algorithms on the 5 benchmark matrices.
The benchmark matrices are listed in Table~\ref{table:matrices}.

\begin{table}
\begin{center}
\begin{tabular}{|l|rrrrr|}
\hline
Algorithm :    &   \multicolumn{5}{c|}{graph, with run time in seconds}  \\
 package       &   Kron    &   Urand   &   Twitter  &  Web    &    Road  \\
\hline
BC   : GAP     &  31.52    &  46.36    &  10.82     &  3.01   &    1.50  \\
BC   : SS      &  23.61    &  32.69    &   9.25     &  8.20   &   34.40  \\     %
\hline
BFS  : GAP     &    .31    &    .58    &    .22     &   .34   &     .25  \\
BFS  : SS      &    .52    &   1.22    &    .33     &   .66   &    3.32  \\     %
\hline
PR   : GAP     &  19.81    &  25.29    &  15.16     &  5.13   &    1.01  \\
PR   : SS      &  22.17    &  27.71    &  17.21     &  9.30   &    1.34  \\     %
\hline
CC   : GAP     &    .53    &   1.66    &    .23     &   .22   &     .05  \\
CC   : SS      &   3.36    &   4.47    &   1.47     &  1.97   &     .98  \\     %
\hline
SSSP : GAP     &   4.91    &   7.23    &   2.02     &   .81   &     .21  \\
SSSP : SS      &  17.37    &  25.54    &   8.54     &  9.61   &   46.79  \\     %
\hline
TC   : GAP     & 374.08    &  21.83    &  79.58     & 22.18   &     .03  \\
TC   : SS      & 917.99    &  34.01    & 239.58     & 34.65   &     .23  \\     %
\hline
\end{tabular}
\caption{Run time of GAP and LAGraph+SS:GrB
\vspace{-2.5ex}
\label{table:results}}
\end{center}
\end{table}

\begin{table}
\begin{center}
\begin{tabular}{|l|rrr|}
\hline
graph   & nodes        & entries in $A$ & graph kind \\
\hline
Kron    & 134,217,726 &  4,223,264,644 &  undirected   \\
Urand   & 134,217,728 &  4,294,966,740 &  undirected   \\
Twitter &  61,578,415 &  1,468,364,884 &  directed     \\
Web     &  50,636,151 &  1,930,292,948 &  directed     \\
Road    &  23,947,347 &     57,708,624 &  directed     \\
\hline
\end{tabular}
\caption{Benchmark matrices\label{table:matrices}
(\url{https://sparse.tamu.edu/GAP})}
\vspace{-2.5ex}
\end{center}
\end{table}

With the addition of the bitmap (needed for the pull step), the
push/pull optimization in BC resulted in a nearly 2x performance gain in the
GraphBLAS method for the largest matrices, as compared to the SS:GrB version
used for the results presented in \cite{DBLP:conf/iiswc/AzadABBCDDDDFGG20}.

With this change, the BC method in LAGraph+SS:GrB is not only expressible in a
simple, elegant code, but it is also faster than the highly-tuned GAP benchmark
method, \verb'bc.cc', for the three largest matrices (1.3x for Kron, 1.5x for
Urand, and 1.2x for Twitter).

The bitmap format (which makes push/pull optimization
simple to express, and fast) and the $\grbanysecondi$ %
semiring, the BFS of a
directed or undirected graph is easily expressed in GraphBLAS, and has a
performance that is only about 1.5x to 2x slower than the GAP benchmark.  We expect
the remaining performance gap arises from two issues:

\begin{enumerate}
\item
GAP assumes that the graph has fewer than $2^{32}$ nodes and edges, and
thus uses 32-bit integers throughout.  GraphBLAS is written for larger
problems, and thus relies solely on 64-bit integers.  This cannot be easily
changed in GraphBLAS, but rather than ``fixing'' GraphBLAS to use smaller
integers, the GAP benchmark suite should be updated for larger
graphs.  In the current GAP benchmark graphs, two graphs are
chosen with almost exactly 4 billion edges.  Graphs of current interest in
large data science can easily exceed $2^{32}$ nodes and edges \cite{9286235}.

\item In GraphBLAS, the BFS must be expressed as two calls.  The first computes
$\grbv{q} \grbmask{\grbneg \grbv{p}} = \grbv{q}\grbt \grbm{A}$, and the second updates the parent vector,
$\grbv{p} \grbmask{\grbstr{\grbv{q}}} = \grbv{q}$:

{\footnotesize
\begin{verbatim}
  GrB_vxm (q, p, NULL, semiring, q, A, GrB_DESC_RSC) ;
  GrB_assign (p, q, NULL, q, GrB_ALL, n, GrB_DESC_S) ; \end{verbatim}}

In GAP's \verb'bfs.cc', these two steps are fused, and the
matrix-vector multiplication can write its result directly into the parent vector
\verb'p'.  This could be implemented in a future GraphBLAS library, since the
GraphBLAS API allows for a non-blocking mode where work is queued and done
later, thus enabling a fusion of these two steps.  SS:GrB exploits the
non-blocking mode (for its lazy sort, pending tuples, and zombies) but does not
{\em yet} exploit the fusion of \verb'GrB_vxm' and \verb'GrB_assign'.  We
intend to exploit this in the future.
\end{enumerate}

Note that for the Road graph,
LAGraph+SS:GrB is quite slow for all but PageRank (PR).
The primary reason for this is the high diameter of the Road graph
(about 6980).  This requires 6980 iterations of GraphBLAS in the BFS, each with
a tiny amount of work.  Each call to GraphBLAS does several \verb'malloc' and
\verb'free's, and in some cases the workspace must be initialized.  A future
version of SS:GrB is planned that will eliminate this work entirely, by
implementing an internal memory pool.  There may be other overheads, but we
hope that a memory pool, fusion to fully exploit non-blocking mode, and other
optimizations will address this large performance gap for the Road graph for
these algorithms.

LAGraph+SS:GrB is also up to 3x slower than the GAP for the triangle counting
problem (for all but the Road graph, where it is even slower).  This
performance gap can be eliminated entirely in the future, if the \verb'GrB_mxm'
and \verb'GrB_reduce' are combined in a single fused step, by a full
exploitation of the GraphBLAS non-blocking mode.  The current method computes
$\grbm{C} \grbmask{\grbstr{\grbm{L}}} = \grbm{L}\grbm{U}\grbt$, followed by the reduction of $\grbm{C}$ to a single
scalar.  The matrix $\grbm{C}$ is then discarded.  All that GraphBLAS needs is a fused
kernel that does not explicitly instantiate the temporary matrix $\grbm{C}$.
This is permitted by the GraphBLAS C API Specification, but not yet implemented
in SS:GrB.

\section{Conclusion}
\label{sec:conclusion}

In this paper we introduced the LAGraph library, the rationale behind its design,  
and a performance baseline with the GAP benchmark suite.   We also introduced
a notation for graph algorithms expressed in terms of linear algebra which we hope becomes
a consensus-notation adopted by the 
larger ``Graphs as Linear Algebra'' community.

This paper defines the foundation for our future work on the LAGraph project.  
We plan to explore Python wrappers for LAGraph that work well for data analytics workflows.  
In addition to the GAP benchmark, which focuses on graph algorithms, we will  
investigate end-to-end workflows based on the LDBC Graphalytics benchmark~\cite{DBLP:journals/pvldb/IosupHNHPMCCSAT16}.

Algorithmically we see a number of research directions to pursue.   With end-to-end workflows, the performance
of data ingestion heavily impacts performance.  We are interested in improving data ingestion performance
by exploiting a CPU's SIMD instructions~\cite{DBLP:journals/vldb/LangdaleL19}.  We are also interested in how  
LAGraph maps onto GPUs using versions of the GraphBLAS optimized for GPUs.

\section*{Acknowledgements}

G.~Sz\'arnyas was supported by the SQIREL-GRAPHS NWO project.
D.~Bader was supported in part by NSF CCF-2109988 and NVIDIA (NVAIL Award).
T.~Davis was supported by NSF CNS-1514406, NVIDIA, Intel, MIT Lincoln Lab,
Redis Labs, and IBM.
This material is also based upon work funded and supported by the Department of
Defense under Contract No.~FA8702-15-D-0002 with Carnegie Mellon University for
the operation of the Software Engineering Institute, a federally funded research
and development center [DM21-0298].
References herein to any specific commercial product, process, or service by trade name, trade mark, manufacturer, or otherwise, does not necessarily constitute or imply its endorsement, recommendation, or favoring by Carnegie Mellon University or its Software Engineering Institute.

\bibliographystyle{IEEEtranS}
\bibliography{ms}

\end{document}